# Photoconductive receivers at 1030 nm for high average power pulsed THz detection

Tim Vogel, Samira Mansourzadeh, Uttam Nandi, Justin Norman, Sascha Preu, *Member, IEEE*, and Clara J. Saraceno, *Member, IEEE*

*Abstract*— In the last few years, many advances have been made in the demonstration of high-average power pulsed THz sources; however, little effort has been made to study compatible sensitive field-resolved detectors. Here, we investigate ErAs:InAlGaAs photoconductive receivers optimized for a probe wavelength of 1030 nm and thus suitable for the new class of high-power ultrafast Ytterbium-based laser sources for THz generation and detection. The performance of the receiver is tested with a few-cycle THz source with high average power up to 20 mW and the dynamic range and saturation behavior of the receiver is thoroughly characterized. Under optimized settings, a dynamic range of more than 115 dB is reached in a 120 s measurement time with 20 mW of THz average power, which is among the highest reported values to date. By reviewing the state-of-the art in TDS measurement and post-processing technology, we identify current limitations and guidelines for further increasing the dynamic range towards 150 dB in short measurement times using high average power THz systems.

## I. INTRODUCTION

THz-Time Domain Spectroscopy (THz-TDS) is a widely used tool in fundamental science but also in applied research and industry. In scientific research, it is mostly deployed to study low-energy dynamics in condensed matter using time-resolved spectroscopy. Some recent examples include the measurements of THz dynamics in glycol-water systems [1], ultrafast dynamics in topological antiferromagnets [2], or the non-linear conductivity response of graphene [3]. In addition to their use in fundamental studies, THz-TDS is increasingly used in industrial settings, where it is most popular for quality control, for example for non-contact thickness gauging of sub-wavelengths coatings [4] or water-proof textiles [5].

The wide adoption of THz-TDS is largely due to the significantly higher sensitivity of the corresponding field-resolved detection compared to well-established Fourier-transform infrared spectrometer (FTIR) methods. Additionally, access to the full field provides access to the phase information of the THz pulses, thus allowing to extract the full complex dielectric function of samples at high accuracy [6]–[8]. Nevertheless, in most applications where THz-TDS acts as a weak, broadband probe, often signal-to-noise ratio and/or dynamic range (DR) are critically sought after (the used definitions for SNR and DR are presented in the following section).

We focus our attention here on reaching highest dynamic range in the frequency domain from THz-TDS – which is most commonly required in linear spectroscopy applications of TDS. Typically, photoconductive emitters and receivers are the most popular choices for linear (i.e. low THz field strength) high-DR, broadband THz-TDS measurements, as they offer highest detector sensitivity and conversion efficiencies from the driving NIR pulses to the THz. Most commercial systems are based on this technology, typically combining compact, low-noise 1550 nm fiber laser technology with highly efficient photoconductive emitters and highly sensitive receivers. Commercial systems based on an oscillating delay line deliver dynamic ranges on the order of 93 dB in the frequency domain in a measurement time of 150 s, and bandwidths of up to 5.6 THz [9]. Yahyapour et al. showed an impressive speed increase using ECOPS (electronically controlled optical sampling) [10], which uses the repetition rate difference between two mode-locked lasers and modulates one of the repetition rates, which introduces a delay and makes a mechanical delay line obsolete. 60 dB of DR are reached in less than 1 s due to the high acquisition rate (1600 traces/s), but a dark trace to verify the flat noise floor is missing as well as the behavior for longer scan times. Kohlhaas et al. achieved record dynamic ranges up to 111 dB with a specialized rhodium-doped photoconductive receiver in a measurement time of 120 s [11]. Nandi et al. have achieved 110 dB in a measurement time of about 60 s (1000 traces) using ErAs:In(Al)GaAs photoconductors and a source power of 0.47 mW [12]. Recently, Kohlhaas et al. also showed 0.97 mW of THz average power from a photoconductive emitter. The average of 1000 traces resulted in 101 dB DR and more than 6 THz of bandwidth, but no total duration for the acquisition time was given. Yardimci et al. implemented a plasmonic nanocavity

We acknowledge funding by the Deutsche Forschungsgemeinschaft (DFG) of the SFB/TRR196 MARIE project M01 and C07 and the DFG project PR1413/3-2. Funded by the Deutsche Forschungsgemeinschaft (DFG, German Research Foundation) under Germany´s Excellence Strategy – EXC-2033 – Projektnummer 390677874 - Resolv. The project "terahertz.NRW" receives funding from the programme "Netzwerke 2021", an initiative of the Ministry of Culture and Science of the State of Northrhine Westphalia. The sole responsibility for the content of this publication lies with the authors.

*(Corresponding author: Tim Vogel, tim.vogel-u81@ruhr-uni-bochum.de).*
Tim Vogel, Samira Mansourzadeh, and Clara J. Saraceno are with the Photonics and Ultrafast Laser Science (PULS) group at the Ruhr-University Bochum, Germany.
Justin Norman is with Quintessent Inc., Santa Barbara, USA.
Uttam Nandi and Sascha Preu are with the Technical University Darmstadt, Germany.
The dataset can be found under https://doi.org/10.5281/zenodo.8200474.

to increase the responsivity of a photoconductive receiver and reached over 100 dB of DR and 6 THz bandwidth with a low probe power of 0.1 mW, but unfortunately also did not specify the acquisition time [13].

Previous studies aiming at increasing DR have mostly focused on optimizing detector sensitivity and laser noise, whereas one widely unexplored possibility is to increase the average power of the emitter. This is most likely because PC emitters and receiver technology have mostly focused on reducing cost for market; thus, higher average power ultrafast laser drivers were not explored. The highest THz powers reported in table-top, fiber-based high DR systems based on photoconductive antenna is on the order of 1 mW [14]. The highest reported THz power using standard photoconductors (non-plasmonic) driven by high power ultrafast lasers is 1.5 mW at a laser power of 0.8 W at 800 nm [15], though with a repetition rate of only 250 kHz.

In the last few years, however, the area of high-average power THz-TDS has significantly expanded, with several recent demonstrations taking THz average power levels close to the Watt-level. These advances were fueled by fast-paced progress of high-average power Ytterbium (Yb)-based ultrafast laser systems [16]–[18], which have recently started to be adopted to THz generation. So far, these remarkable advancements have been achieved using optical rectification and two-color plasma filamentation. In [19], Kramer et al. showed 144 mW of THz average power with 2 THz bandwidth in room temperature lithium niobate (LN), using a 350 W, 70-fs laser at 1030 nm at 100 kHz repetition rate; in [20], Buldt et al. achieved 640 mW of THz average power spanning over 20 THz with a two-color plasma filament in a noble gas, also generated with a ultrafast laser with its center wavelength of 1030 nm but a repetition rate of 500 kHz and a pump power of 570 W. More recently, using cryogenically cooled LN, 643 mW were reported in a first conference contribution [21] with 400 W of pump power at 1030 nm and 40 kHz. In [22], 66 mW were demonstrated at 13.3 MHz using room temperature LN. In all these results, the main focus was placed on achieving strong-fields at high repetition rates compared to usual 1 kHz systems. Nevertheless, the repetition rates used in these experiments are far below those of table-top fiber-based systems, typically on the order of 100 MHz. For TDS, however, high repetition rates allows us to perform averaging over many pulses per time and are thus of utmost importance to lower the noise floor. This explains why lower power table-top fiber-coupled TDS systems have yet demonstrated the highest DR, despite working with about two to three orders of magnitude lower THz power levels. We note in this context that these advances open the door to upcoming developments in photoconductive emitters excited at high average power as well.

Here, we explore the route of reaching high DR by combining novel low-noise photoconductive receivers designed for 1030 nm based on ErAs:InAlGaAs with recently developed high THz average power sources based on optical rectification (using the tilted-pulse front in lithium niobate scheme). Our high-power THz source, which can deliver a day-to-day THz power of >20 mW, is driven by a home-made ultrafast modelocked laser delivering more than 100 W of average power and 550 fs pulses at 13.4 MHz. The combination of non-linear crystal and photoconductive emitters or receiver were so far rare and focused mostly on improving the detection bandwidth [23]–[25]. We thoroughly characterize the behavior of this novel, low-noise emitter at high THz average power and with 1030 nm probe wavelength from our high-power laser. As a part of this exploration, we maximize the dynamic range in frequency domain by sweeping THz power and laser probe power allowing us to gain insight on the saturation behavior and data-processing required to reach a high dynamic range above 110 dB in our high-power operation regime. For such extreme DRs, all system aspects, not only source and receiver performance have to be optimized. In this regard, we review the state of the art and identify important steps to achieve DRs beyond the yet achieved 111 dB. Under best conditions, we reach 115 dB dynamic range in a 120 s measurement time, which is comparable with current state-of-the-art values with lower excitation powers. We identify and the current limitations in our system and discuss future options to reach dynamic ranges of 150 dB for a reasonable measurement time, when we make full use of the average power available.

## II. BASIC CONSIDERATIONS AND DEFINITIONS USED

Since the main goal of our paper is discussing DR in high-power TDS systems, we start by summarizing some definitions that are used throughout this article. We note that these definitions can be found in extensive multiple pieces of literature; however, our aim here is to concisely summarize the definitions here. We also note that discrepancies in the definitions are often seen in the literature, therefore it is important to clarify the definitions we employ here. Concerning signal-to-noise ratio (SNR) and dynamic range (DR), we use the common definition as introduced by Naftaly and Dudley for the time domain [26]:

$$\text{SNR} = \frac{\text{mean magnitude of amplitude}}{\text{standard deviation of amplitude}} \quad (1)$$

$$\text{DR} = \frac{\text{maximum magnitude of amplitude}}{\text{rms of the noise floor}} \quad (2)$$

Where SNR is a metric to describe the minimal detectable signal change and rms stands for root-mean-square. In ref. [26], the amplitude at time delays before the arrival of the THz pulse is used as an approximation of a the noise floor. As it was discussed by Skoromets et al. [27], this can lead to a biased underestimate of the true noise floor due to correlated noise when the THz pulse is present. A more precise determination of the dynamic range in time domain can be obtained, when a separate measurement with blocked THz beam is acquired, in this paper called "dark" measurement. The typically used definition by Naftaly and Dudley for SNR and DR in the frequency domain is not as clearly defined [26]:

$$\text{SNR} = \frac{\text{mean FT amplitude}}{\text{standard deviation of FT}} \quad (3)$$

$$\text{DR} = \frac{\text{mean FT amplitude}}{\text{noise floor}}, \quad (4)$$

where the time domain-based traces are Fourier-transformed (FT) to the frequency domain. Naftaly and Dudley approximate the noise with the (constant) amplitude at higher THz frequencies, which are above the bandwidth of the detected THz radiation. This constant value serves, approximately, as the level of the noise floor. The noise floor for real world measurements can be frequency-dependent and especially low THz frequencies suffer from increased noise or even a deterministic signal arising from rectification of the optical probe pulse due to imperfect contacts or non-symmetric illumination of the receiver, which often leads to misleading values of the DR. Thanks to a separate "dark" measurement, we can frequency-resolve the noise floor of the system affected by noise sources such as electronic noise, beam pointing and instabilities of the pump-probe length difference. Single cycle THz sources typically contain frequencies close to 0 THz, which are particularly affected by the increased noise at low frequencies. A more precise evaluation of the dynamic range requires us to consider the noise floor within the bandwidth of the detected THz radiation. Especially for small delay ranges around the single cycle pulse, the resolution in the frequency domain is limited and a simple mean of the noise floor can be affected by outliers. In our measurements, we take the median of the "dark" amplitudes in the THz detection bandwidth, which is more robust against outliers. Additionally, any processing steps done to the THz data, if applied in the same way to the "dark"-data, make this approach more robust against errors in assessing the bandwidth or dynamic range of the THz-TDS.

The peak dynamic range is often implicitly communicated as "dynamic range" as a single number, even though it is highly frequency dependent. We note however, that each application scenario may have different criteria of relevance, and DR is only one metric to characterize the sensitivity of a THz measurement.

A multiplicity of factors affects the signal quality, and thus the SNR and DR, of a THz-TDS. An extensive overview of random and systematic errors in THz-TDS was done by Withayachumnankul et al. [28], [29]. The most common uncertainties stem from laser intensity fluctuations, laser beam pointing, noise of the receiver and delay line registration errors. Readers may find multiple relevant literature on how these and other parameters influence the DR, which strongly depends on the type of emitters and receivers used [30]–[35]. The model developed by Duvillaret et al. [36] allows to understand the importance of the relative intensity noise (RIN) of the laser source and SNR of the THz-TDS,

$$\text{SNR} = \frac{P_{\text{opt}}}{\sqrt{\text{RIN} \cdot P_{\text{opt}}^2 + C}} \quad (5)$$

With $P_{\text{opt}}$, the optical power and $C$, a weak model parameter, which is independent of the THz signal and results from white noise in the receiver. Achieving a high DR due to averaging can only be achieved when the SNR is sufficiently high. It should be noted, that this model was developed for photoconductive emitter and receiver pairs and other publications do not see a strong limitation on the DR by laser fluctuations [37].

A more recent study by Mohtashemi et al. [38] takes into account the non-Gaussian nature of the noise sources, arising from the inherent Fourier transformation of the time-domain data. Taking these effects into account, a less biased estimate of the uncertainty of retrieved parameters like the refractive index can be achieved. The group of Prof. Peretti is currently developing a sophisticated software package called Correct@TDS and is incorporating this advanced model to characterize and correct for some of the noise terms and get a more reliable and comparable definition of DR [39]. In this paper though, we stay close to the DR definition from Naftaly and Dudley [26], to keep the results comparable to the current state-of-the-art results.

## II. METHODS

### II.1 EXPERIMENTAL SETUP

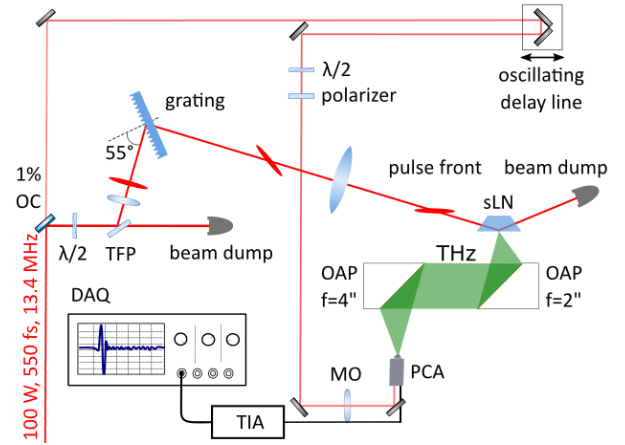

**Fig. 1** Setup for THz generation with stochiometric lithium niobate (sLN) with the tilted pulse front approach at 1030 nm. Coherent detection of THz radiation is done in the photoconductive antenna (PCA) receiver, where a microscope objective (MO) is focusing the probe beam as an optical gate from the backside. The pump and probe beam power are adjusted with a thin-film polarizer (TFP) or a polarizer in connection with a $\lambda/2$-waveplate. The signal is amplified in a transimpedance amplifier (TIA) and captured with a data acquisition (DAQ) device.

The receiver material is tailored to an Yb-based laser system as driver, thus making it suitable for modern high-power ultrafast laser systems: it is composed of 90 periods of 1.6 monolayers p-delta-doped ErAs and 18 nm of a digital alloy of $(\text{In}_{0.53}\text{Ga}_{0.47}\text{As})_{0.46}(\text{In}_{0.52}\text{Al}_{0.48}\text{As})_{0.54}$. The material was grown lattice-matched on semi-insulating Fe:InP substrates and capped with 5 nm InGaAs for protection against oxidation. The targeted band gap of the InAlGaAs host lattice is 1.07 eV, well-suited for operation at the laser wavelength of 1030 nm (1.204 eV). The p-delta-doped ErAs precipitates ensure low carrier lifetime at high resistance. Details on the material system can be found in [12] and [40]. An H-dipole antenna with a center dipole length of 25 µm and a photoconductive gap of 5 µm connects to the photoconductor.

As a driving laser for all the experiments presented here, we use a home-built modelocked thin-disk oscillator at 1030 nm, that generates up to 100 W of average power and 550 fs pulses at 13.4 MHz repetition rate, corresponding to a pulse energy of 7.5 µJ and a peak power of 12 MW. The oscillator is a semiconductor saturable absorber (SESAM) soliton-modelocked laser [41], [42]. We do not focus here on the details of the laser system, which have been presented elsewhere [43].

An overview of the THz setup is shown in Fig. 1 and more details about the THz source can be found in [22]. The pulse train from the laser is split in pump and probe with a 1% output coupler (OC). 99% of the beam is guided through a lambda/2-waveplate in a motorized rotation stage (PI DT-80) followed by a thin-film polarizer (TFP), allowing to precisely control the pump power. The reflected beam after the TFP is focused by a lens and a transmission grating (1600 lines/mm), which introduces a pulse front tilt, as introduced by Hebling et al. [44].

The beam is imaged from the grating, into the trapezoidal-shaped stoichiometric lithium niobate crystal (sLN), where the pump beam generates efficiently THz radiation. After the main pass on the crystal where THz is generated, the pump beam is totally internally reflected and captured by a water-cooled beam dump. The THz beam diverges strongly and is collected by a 50.8 mm diameter off-axis parabolic (OAP) mirror with a focal length of 50.8 mm. The collimated beam is then focused with a 50.8 mm OAP with a focal length of 101.6 mm onto the ErAs:InAlGaAs photoconductive antenna (PCA) receiver. The probe beam, originating from the transmission of the 1% OC, is guided over a fast-oscillating delay line (APE ScanDelay, 15 ps) and a lambda/2-waveplate in another motorized rotation mount (PI DT-80), which allows in connection with a polarizer to control the probe laser power, too. Due to the small gap of the PCA of 5 µm, strong focusing is required. A suitable microscope objective for the near-infrared range in employed (Mitutoyo M Plan Apo NIR 20X). The rectified current generated by the PCA, which is proportional to the THz electric field, is amplified by a transimpedance amplifier (TIA, Femto DLPCA-200) with a gain of $10^7$ V/A and read out by either by an oscilloscope (for real time alignment) or by data acquisition hardware (DAQ, Dewesoft Sirius Mini). The rotation angle of the lambda/2-waveplate in the pump-arm is calibrated to a THz power meter (Ophir 3A-P-THz) and, respectively, the rotation angle of the lambda/2-waveplate in the probe-arm to a power meter suitable for 1030 nm (Thorlabs S401C). The THz power and laser probe power can thus be precisely controlled and selected in a systematic way.

The delay line consists of a shaker with up to 20 Hz modulation frequency. At a measurement time of 120 s up to 4800 THz traces are thus captured for each combination of THz pump power and laser probe power. As Withayachumnankul and Naftaly pointed out [29], there is on the one hand the need to average due to random errors, which increases the dynamic range, but on the other hand, long-term drifts and environmental changes can prohibit a reduction of noise when the number of traces to average is increased and limit, or even deteriorate, the dynamic range. With this in mind, the fast-oscillating delay line "freezes" environmental factors to a much higher degree than the classic "step-and-settle" approach and a reduction of uncertainty is achieved for the whole trace instead of a point-by-point basis. In our case, the collected traces over the period of 120 s are averaged in time domain. The "step-and-settle" approach, meaning the delay stage is moved one step and waits until the lock-in amplifier is settled and a quasi-DC value is read out, is mainly used due to the low repetition rate of amplifier lasers, which typically offer repetition rates ≤ 1 kHz. Besides the extra hardware like the lock-in amplifier for the "step-and-settle"-approach, the pump or THz needs to be modulated, effectively loosing 50% of its power. With our MHz repetition rate this setup is limited by the shaking rate of the oscillating delay line. Even if we could increase the speed of the oscillating delay line, the bandwidth of our transimpedance amplifier (max. 500 kHz) would be in theory a stricter limitation to the maximum shaking rate than the number of pulses.

### II.2 NOISE FLOOR AND POST-PROCESSING OF THz TRACES

Reaching a high dynamic range requires very careful optimization of the setup, precise knowledge of the different sources of noise, and careful data processing. In order to understand the limitations of our setup and push the DR to high values, we show how typical factors originating from the system components (shaker pointing, laser power fluctuations) or from the data processing (strong low-pass filtering) can result in inaccuracies when characterizing the DR. Whereas many of these aspects have been discussed in the literature, this remains often neglected in recent reports, therefore we aim here to give an illustration of the effects and their importance, as well as clearly explain how our data is processed.

First, we focus our attention on the noise floor level, which is critical data that needs to be characterized for a proper assessment of the DR. For each tested laser probe power one dataset is captured with the THz beam blocked. This "dark" trace captures the electronic noise floor as well as the influence of beam pointing from imperfect alignment of the oscillating delay line. The latter causes a change of resistivity of the photoconductor and thus modulates the dark trace in a deterministic and repeatable way, which can be, at a later stage, corrected. The data is captured and processed with the same parameters as the THz traces. Acquiring these extra dark traces allows to see the noise floor in a frequency-resolved fashion and identify systematic problems in the detection. When a dark trace is available, one can average the "dark"-power in the bandwidth of the captured THz radiation and does not need to assume a constant noise floor, which is typically extrapolated from higher frequencies in the literature and can result in incorrect DR values. Although most commercial THz-TDS systems use fast oscillating delay lines and make high dynamic ranges in a short measurement time possible, data processing of cutting and averaging the single traces needs to be done in a well-defined manner. The following discussion starts with averaged THz/dark trace in time domain.

Fig. 2 illustrates the importance of a proper noise characterization and the necessary post-processing steps to extract the dynamic range, in particular in the specific case of a fast-oscillating delay line which, as we mentioned before, is desirable to make fast THz scans. In Fig. 2 a) b) and c) (first column) the traces are only averaged without post-processing. In Fig 2a), the averaged time trace is shown when the THz beam is blocked (black), with a strong parabolic shape imprinted on the high frequency noise. This originates from the beam pointing effect of the oscillating delay line, which consists of a retroflector in a freely hanging voice coil design, which can only approximate a harmonic oscillator with a single frequency (20 Hz). Additionally, the movement of the shaker shows higher harmonics and residual beam pointing errors. This results in an offset in amplitude for different delay line positions, which are repeatable for each single trace and thus appear in the averaged trace. Fig. 2b) shows the photocurrent when the receiver measures the electric field of the THz radiation. The curved baseline cannot be seen in Fig. 2b), because the generated photocurrent is orders of magnitude higher than the current when the THz light is blocked. The plot in Fig. 2c) shows both curves in the frequency domain, indicating multiple problems: Amplitude fluctuations at higher THz frequencies are smoothened out and the curve (orange) does not reach the noise floor (black). Because our oscillator has rather long pulse durations, our THz spectrum does not exceed 2.5 THz, therefore, the observed roll-off at high frequencies is no real signal. Additionally, the dark trace has, due to the strong parabolic shape in time domain, higher amplitudes for low frequencies. In Fig. 2d)-f) the same data is used, this time only subtracting a 4$^{th}$ order polynomial fit of the dark trace in time domain. Fig. 2d) shows now only high frequency noise, which is approximately equally distributed in time. Even though the same polynomial is also subtracted from the THz time trace, it has a negligible effect, since the photocurrents are much larger than the correction. In the frequency domain, the amplitude of the dark trace at low frequencies is reduced and close to the value of the noise floor at high frequencies. However, the offset between THz and dark trace persists as well as the smooth amplitude roll-off of the THz trace. It is due to the fact that the finite measurement window is de facto a product of an infinite data set (i.e. the true signal) with a rectangular function with a width of the time window. The Fourier transform of the product leads to a convolution of the THz spectrum and the Fourier transform of the rectangular function, i.e. a sinc² function whose envelope rolls off as $f^{-2}$. For sufficiently long time traces, the zeroes of the sinc² function would become visible. In the present manuscript, however, the time trace is just 15 ps long such that the zeroes are out of range in the frequency domain. To mitigate this effect, the next post-processing step consists of the application of a smooth windowing function to the time domain data (Fig. 2 g)). We chose a 10$^{th}$-order, super-gaussian window with a smooth roll-off,

$$SG(\sigma, \tau, k, n) = \exp\left(-2^n \cdot \ln(2) \cdot \left|\frac{\tau - \frac{k-1}{2}}{\sigma}\right|^n\right), \quad (6)$$

where $\sigma$ is the window width in percent, $\tau$, the index of the data from 0 to the length of the time domain signal, $k$ the length of the time domain data in samples, and $n$ the window order.

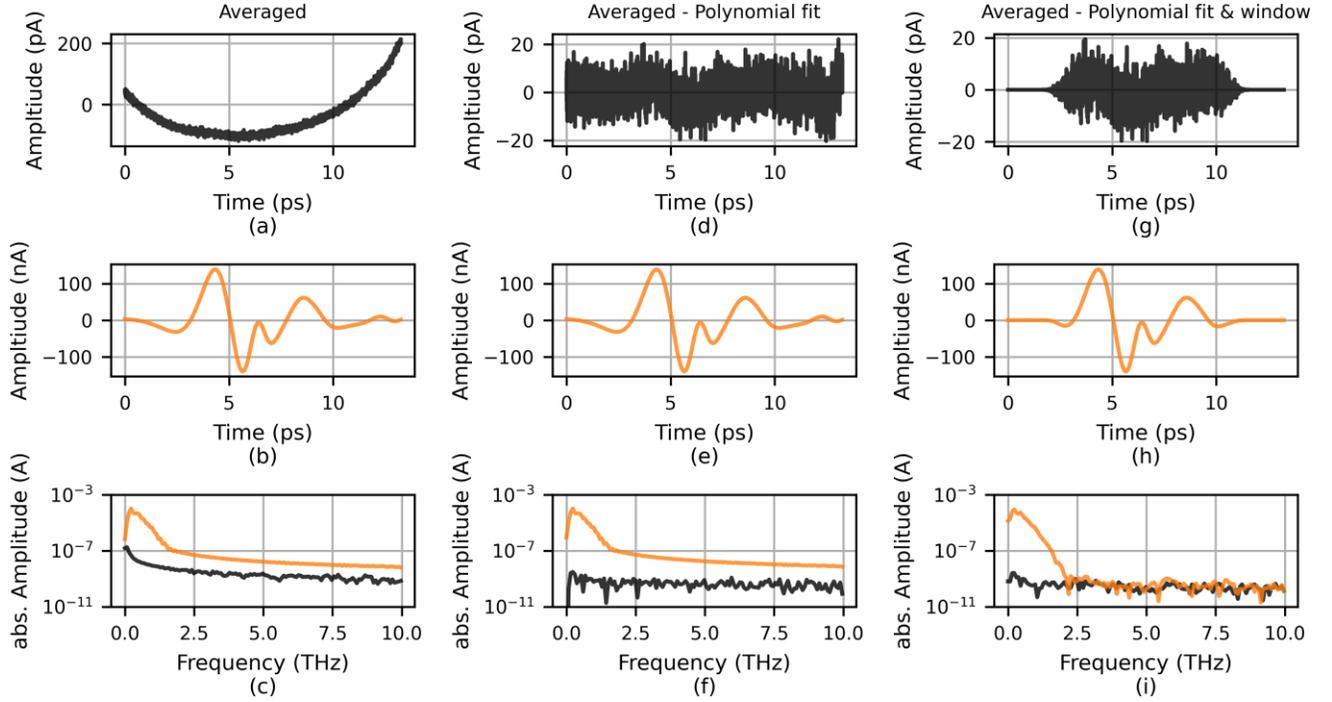

**Fig. 2** Noise traces are acquired when the THz beam is blocked and are displayed in black. THz traces are displayed in orange. Only averages are displayed in the left column (a) to (c) from averaged dark trace and averaged THz trace in time- and frequency-domain. Note the different scale of dark and THz trace (pA vs. nA). A 4$^{th}$ order polynomial is fitted to the dark trace, since the low frequency component is a reproducible artifact of the non-ideal movement of the oscillating delay line. After subtraction of this polynomial, the time-domain of the dark trace shows only white noise components (d). The same polynomial is subtracted from the THz trace, but shows miniscule effect in time-domain (e). The frequency domain (f) shows a reduction in amplitude for low frequencies to a similar level like higher frequency components. Additionally, windowing is applied to the time-domain which limits artifacts introduced by the FFT due to the finite size of the time window, bring start and end of the dark- (g) and THz-trace (i) in time-domain smoothly to zero. This shows the real noise floor of the THz trace in frequency-domain (i) and makes the bandwidth at 2.2 THz visible.

A study by Vázquez-Cabo et al. showed the effect of windowing of THz-TDS data in more detail and how it improves the dynamic range and accuracy of the extracted parameters like refractive index and absorption coefficient [45]. We can observe the effect of the windowing in Fig. 2g) – i). The dark trace is smoothly tapered off to zero (Fig. 2g)) and the THz trace is treated the same way (Fig. 2h)). The frequency domain in Fig. 2i) shows not only the increased dynamic range, but also the overlap between dark trace and THz trace in the high frequency limit. This allows us to determine properly the bandwidth of the lithium niobate-based THz source to be approximately 2.2 THz, in excellent agreement with the inverse pump pulse duration of 550 fs of our oscillator.

We note that there is currently no consensus how high dynamic range values should be reported, making fair comparisons challenging. Industrial guidelines have been developed in order to make a comparison between different manufacturers and systems possible [46], however, this standard is typically not followed by researchers. Results of DR are presented with different time windows, different source bandwidths and different post-processing steps and often these important values are not given in literature. Given the critical importance of these steps in the performance of the TDS, we would like to encourage future reports to provide details about data processing of the traces to enable comparisons of different systems. Another important factor is the nature of the delay stage used: the total delay range needed depends on the experimental question to be solved. For retrieving broad absorption features or characterizing the general THz source bandwidth, a short delay range like used in this paper is sufficient. High resolution THz spectroscopy needs longer delay range for smaller frequency steps, but comes at the cost of reducing the dynamic range, as it was shown by Mickan et al. [47]. In this case, the different effects affecting the measurement need to be studied separately.

In spite of this difficulty in properly comparing DR values, we aim here to benchmark our setup in terms of DR to existing literature, thus we chose the same total acquisition time of 120 s as reported in the recent record dynamic range from Kohlhaas et al. [11]. In a previous conference submission, we used another receiver and reported, without a detailed parameter study, 80 dB in a 10 s measurement time [48]–[50], however this receiver was damaged due to its sensitivity to electrostatic discharges. In this follow-up detailed study, we use a second-generation receiver with a Zener-diode-based protection circuit to mitigate the danger of electrostatic discharge.

## II.3 RESOLUTION OF THZ SPECTRUM

Due to the short delay range of approximately 15 ps and applied windowing in the above results, the frequency resolution is too low to make water vapor absorption lines visible using the shaker described above, which is an important sanity check. The oscillating delay line is therefore exchanged only for this subsection with a delay line with longer range of 300 ps (SMAC LCS50-050). Instead of a free-swinging voice-coil design, a retroreflector is guided with ball-bearings and driven by a voice coil along a rail to travel this longer distance. The emitter and receiver, as well as the data acquisition remain identical.

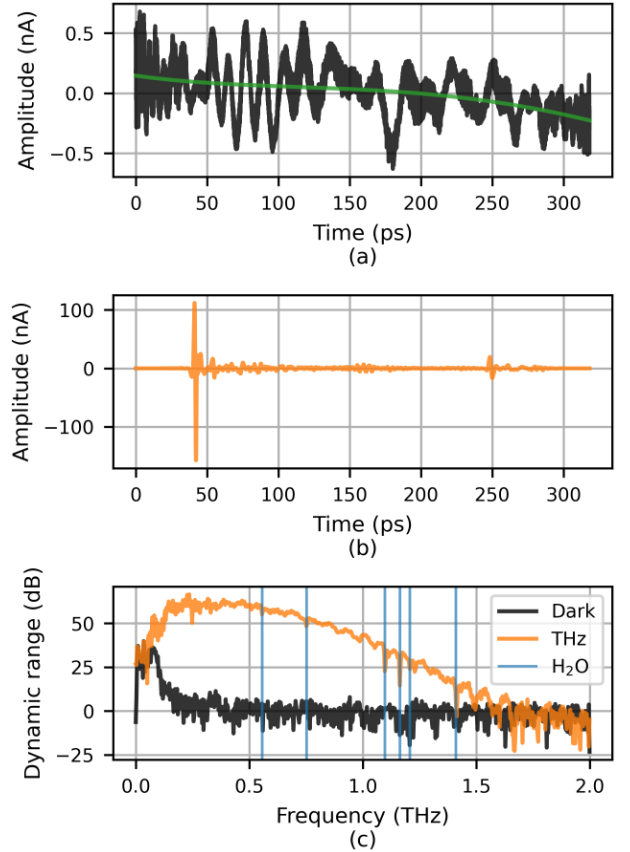

**Fig. 3** Measurement with an increased delay range of 300 ps using a different delay line as described in the text. The dark trace (a) shows a strongly modulated pattern in the averaged trace, which a simple polynomial subtraction cannot compensate. (b) shows the THz recording in time-domain. (c) shows the corresponding spectra of the THz and dark trace, after windowing both traces. The water vapor absorption lines (blue) became visible due to the 20 times higher spectral resolution. The peak dynamic range with this delay line is approximately 60 dB, due to stronger beam pointing and lower number of traces in 120 s.

Unfortunately, the beam pointing stability from this longer delay line is worse and results in strong modulations of the dark trace as shown in Fig. 3a). Thus, a polynomial fit is insufficient to describe its behavior. Fig. 3b) presents the received photocurrent with incident near single cycle THz radiation from

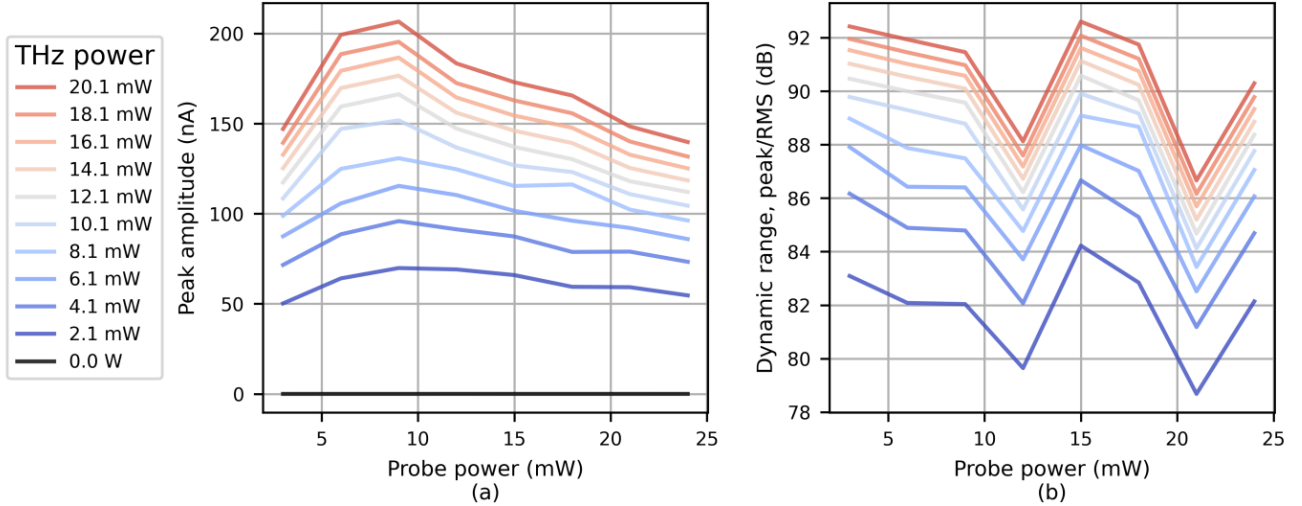

**Fig. 4** (a) shows the peak amplitude in time domain for all measurement conditions. The maximum peak amplitude is 207 nA at 9 mW probe power and max. THz power. The time domain dynamic range is calculated by dividing the peak amplitude of the THz trace by the RMS of the dark trace at the same probe power. Besides two small dips, the dynamic range in the time domain is approximately constant for a given THz power.

the lithium niobate crystal. The ringing after the main peak originates from narrow water vapor absorption lines. A reflection is visible approximately 210 ps after the main THz pulse, originating from the partly reflected THz pulse inside the trapezoidal LN crystal. In Fig. 3c), the dark trace shows a more pronounced increase at low frequencies, owing to the strong modulations of the dark trace. The dynamic range is approximately 60 dB and leads also to a relatively small detected bandwidth.

There are several reasons for this limited dynamic range: The increased noise in the dark trace, originating from beam pointing fluctuations from the longer delay line were already mentioned. Another reason is, that the main information is contained in the time window around the single-cycle THz pulse. From a perspective of accumulated energy along the delay time, the signal is quickly saturated after the main pulse and barely increases, whereas the accumulated noise is continuously increasing [47]. This reduces the dynamic range in frequency domain, when long delay ranges are used. A third factor lies in the accuracy of the delay line position. The short oscillating delay line has an integrated interferometer and gives access to much more accurate position data. The long delay range uses an optical encoder with a repeatability of $\pm 1$ μm. Averaging multiple traces with a larger error in the delay axis leads to a reduction in dynamic range [34]. Finally, since the same total time of 120 s is used as with the short delay range, due to the longer range and slower speed, the total number of acquired traces decreased to approximately 585 as compared to the 15 ps delay line with 4800 traces. The lower number of traces results in a drop of 9 dB in dynamic range.

Nevertheless, the water vapor absorption lines (blue, taken from HITRAN [51], [52]) excellently match the absorption dips in the THz spectrum confirming the attained bandwidth.

### III. RESULTS

#### III.1 INFLUENCE OF THz POWER AND PROBE POWER ON DR

First, we explore the influence of THz power and probe-power in the time and frequency domain. In Fig. 4, we present the results of peak THz amplitude and DR in the time domain for each THz power between 0 mW (dark) up to 20.1 mW, while the laser probe power is varied between 3 mW up to 24 mW. In the following section, this dataset is displayed from different perspectives to gain insights into the behavior of the receiver. At each unique combination of THz power and laser probe power, the acquired traces are averaged and post-processed as described in section II.2.

The peak amplitude of the averaged THz trace in time domain for various THz powers (blue to red) is plotted versus probe power in Fig. 4a). More than 200 nA peak current is generated at maximum THz power. There is a monotonous increase with THz power at all tested probe powers, showing that the receiver is not yet in saturation. A slight drop in peak current for all THz powers after applying more than 9 mW laser power can be observed, but there is no strong roll-over visible. Fig. 4b) shows the calculated dynamic range in time domain, taking the division of the peak current when the THz power is applied over root-mean-square of the dark trace (0 mW) for the given probe power.

The maximum dynamic range in time domain is over 92 dB and shows the same monotonous increase with THz power as in Fig. 4a). Only two dips are persistent at 12 mW and 21 mW. This originates not from the THz traces themselves, but from the dark trace taken at these probe powers. Minute changes in the environment or the laser can lead to a slight increase of the RMS value, thus reducing the dynamic range for all tested THz powers at these probe powers. Nevertheless, the dynamic range in time domain stays in the same order of magnitude. The improvement in DR can be attributed to the polynomial

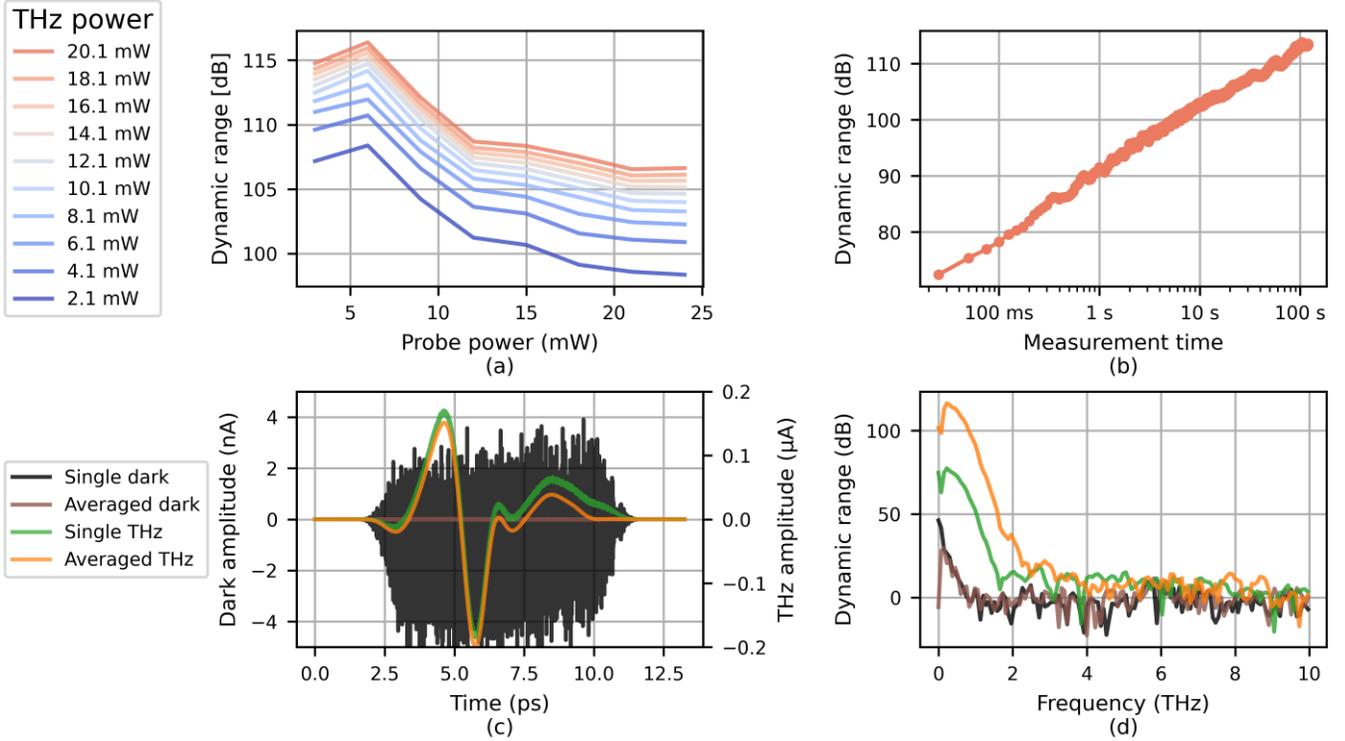

**Fig. 5** (a) shows the dynamic range in frequency domain after 120 s of averaging for various THz power. (b) shows the effect of averaging for the optimal DR with a setting of 20.1 mW of THz power and 6 mW of probe power. The dynamic range increases linearly and does not show saturation. (c) shows for the optimal settings the dark trace for a single trace (black) and averaged over 120 s (brown). There is a substantial reduction of noise due to averaging. On the right y-axis, the amplitude of a single THz trace (green) and averaged over 120 s (orange) is shown. Even though also affected by noise for a single trace, the amplitude barely changed since the signal is stable. Note the different units for the two y-axes. (d) shows the corresponding frequency domain of the beforementioned traces. The median value of the respective dark trace between 0 THz to 2.2 THz is set to 0 dB.

subtraction, as discussed in the methods section. The repeatable deviation in the noise floor, originating from the oscillating delay line, lead to higher RMS values when not corrected. A separate analysis, when the polynomial subtraction was not applied, showed roughly 15 dB lower peak DR in time domain. Fig. 5a) shows the obtained peak dynamic range in the frequency domain. The post-processed data from the time domain is converted to the frequency domain and the peak of the power spectrum is compared to the median value of the noise floor in the observed bandwidth (0 THz to 2.2 THz). We observe again a monotonous increase of DR with THz power, while keeping the laser probe power fixed. In the frequency domain, the peak dynamic range (for all laser probe powers) is obtained at 6 mW probe power. The interplay of the dynamic range in time and frequency domain does not follow a simple analytical relationship [29] and the maximum dynamic range in either domain can be obtained at (slightly) different parameters. The reduced DR at high laser powers may originate from laser noise of the probe pulse combined with slight saturation of the receiver.

We focus now on the optimal performance of our system. First, we compare single shot DR with other existing literature values. In [11], 78 dB of DR for a single trace in 60 ms were achieved. In our case, we achieved 71 dB of DR for a single trace in 25 ms. At first sight, this is surprising : in ref [11], a photoconductive receiver was illuminated with less than 7 pJ of THz pulse energy, but exhibits higher dynamic range (single trace 78 dB) than our emitter with approximately 1500 pJ of THz pulse energy (single trace: 71 dB). We believe this is due to the increased noise level of our high-power laser: the amplitude noise of the laser oscillator is directly coupled from the probe beam into the photoconductive receiver. Our high-power, home-build oscillator is not actively stabilized in power and is pumped with strongly multimode laser diodes, which can result in increased noise levels. As mentioned in the introduction, typical noise levels reported in the literature are 0.1% for high-power, thin-disk lasers (TDL) [53]–[55], to be compared to orders of magnitude lower RIN of 0.005% in [1 Hz, 1 MHz] from low-power, fiber-based laser systems [56]. We note, however, that compared to other high-power laser systems, modelocked TDLs are high-$Q$ resonators, thus can be optimized for very low noise level, even at high power. Besides the amplitude noise of the laser, other noise sources such as beam pointing could be reduced with different means of creating a delay between pump and probe beam or by actively stabilizing the beam. All these improvements would scale the single trace dynamic range to even higher values, offering a straightforward path for reaching record high DR.

In order to measure the improvement of DR via averaging, we average 4800 traces (120 s measurement time as in [11]). Fig. 5b) shows the expected linear relationship of DR versus measurement time. The maximum dynamic range of 115 dB can be reached with 20.1 mW of THz power and 6 mW of laser probe power. We highlight that the pump laser, compared to the typically employed 100-MHz fiber lasers, has nearly one order of magnitude less pulses per time interval but still manages to reach such high dynamic ranges due to the high THz average power. We note that given the trend observed in Fig. 5 b), it should be straightforward to obtain higher DR values by increasing measurement times even further. However, we believe minute-long measurement times are compatible with real-time scenarios, thus we limited ourselves to such durations. We note that 100MHz class thin-disk lasers exist; however prohibitively large average powers would be required to generate THz radiation with comparable conversion efficiency in LN as we demonstrated in [57]. We note also that compared to [11], we have significantly narrower bandwidth due to about 6 times longer laser pulses. Pulse compression techniques could be applied to reach sub-100 fs pulses. This was, however, out of the scope of this manuscript.

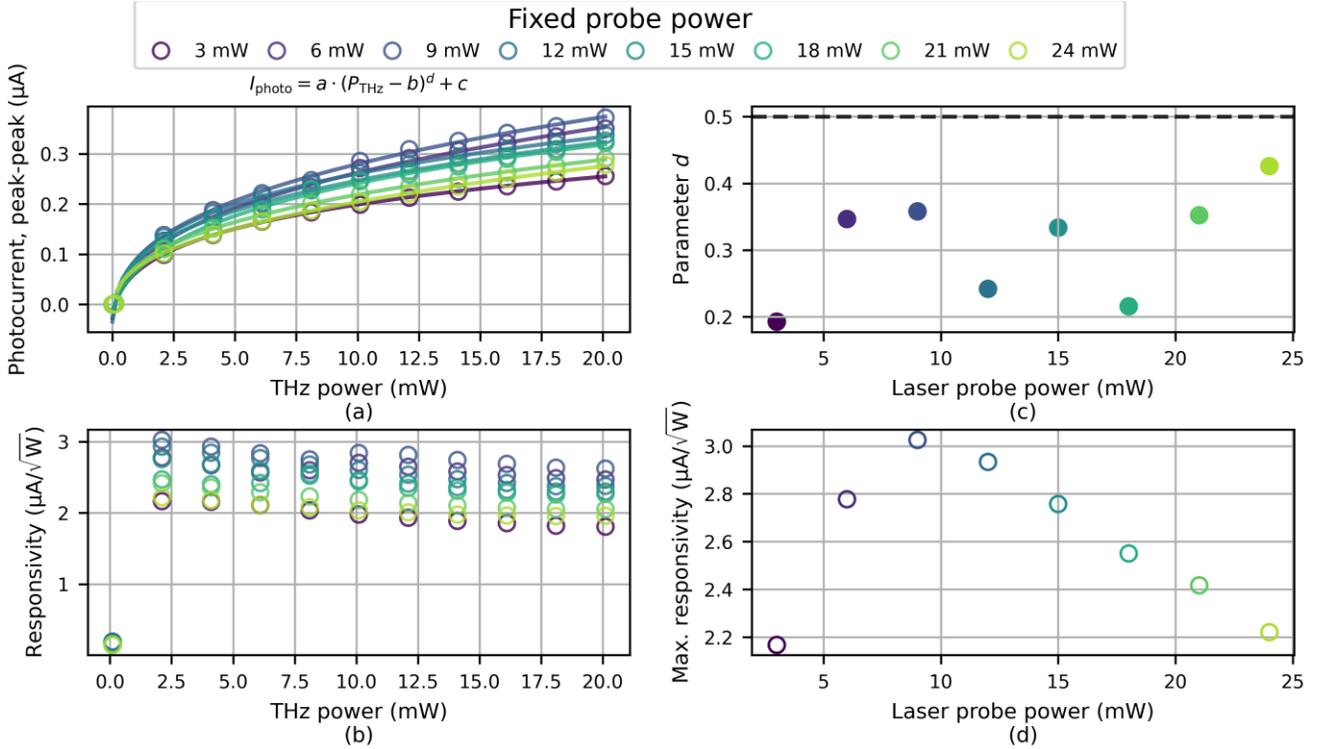

**Fig. 6** Analysis of the saturation behavior of the PCA receiver. (a) Peak-peak photocurrent in time-domain vs. THz power for various probe powers. The open circles are measured data and the continuous line is a curve fit. (b) depicts the THz field responsivity of the receiver, assuming a square root behavior of the THz power. There is no strong downward trend visible, indicating only little saturation. (c) shows the parameter $d$ from the curve fit for the different laser probe powers, where a value of $d = 0.5$ would indicate a perfect receiver with no saturation. The fit-value fluctuates between 0.2 to 0.42, another indication that there is only slight saturation visible. The maximum responsivity is 3.0 µA/$\sqrt{W}$.

Fig. 5c) shows two examples from the acquired dataset in the time domain. The left axis shows the amplitude of the dark trace (no THz) for a single trace (black) and the averaged trace (brown) for the optimized probe power of 6 mW. The noise is similar to white noise and has no imprinted structure. The averaged dark trace is strongly reduced in amplitude due to averaging. On the right y-axis, the amplitude of a single THz trace and the averaged trace is displayed. In an ideal case (besides the additional noise introduced by the THz emitter), both curves should look identical. Systems suffering from low SNR, stemming from amplitude or phase fluctuations, would show a reduced amplitude with increasing number of averages. Compared to the noise, the measured THz electric field should stay constant after averaging. The constant THz value divided by the decreasing power of the noise floor leads to an increase in dynamic range. In our case, both traces have similar structure and amplitude. The measurement range of 120 s seems to be still below any deviation from white noise behavior, allowing to reach higher and higher dynamic ranges with longer measurement time.

Fig. 5d) shows the obtained dynamic range in frequency domain for a single trace (green) and all averaged traces (orange). The dark trace from single (black) and averaged (brown) trace shows a persistent increase of its amplitude at lower frequencies. In future studies, more advanced correction techniques could possibly compensate part of the increased error at low frequencies, if the introduced error is deterministic.

III.2 SATURATION BEHAVIOR

From Fig. 4 we see already that there is no roll-over of the peak electric field in the time domain for higher THz powers, and the dynamic range in time and frequency domain increases monotonically (for fixed laser probe power). This means that the receiver seems to handle more than 20 mW of THz power, which is still larger than state-of-the-art results from photoconductive antenna emitters that reach up to 0.47 mW [12] to 0.97 mW-level [14] and are typically paired with photoconductive receivers. Even record THz powers obtained from photoconductive antenna emitters enhanced with plasmonic nanostructures are limited to 6.7 mW [58] but at this power level do not demonstrate high DR. We note that this was the first-generation of 1030 nm receivers, thus holding great promise for future optimizations.

To gain a more quantitative insight, Fig. 6a) shows the same dataset for various laser probe powers and plotted versus THz power. The photocurrent of an ideal receiver should show a square-root dependence on the THz power. To describe the connection between photocurrent and THz power, we use a simple model:

$$I_{\text{photo}}(a, b, c, d, P_{\text{THz}}) = a \cdot (P_{\text{THz}} - b)^d + c \quad (2)$$

To accustom for a possible systematic error/offset from the power meter, the variable $b$ is introduced. Possible scaling errors and offsets in the measured photocurrent are handled by parameter $a$ and $c$. For the ideal receiver, the parameter $d$ is equal to 0.5. Values below 0.5 indicate saturation. For each fixed probe power, a non-linear least squares algorithm finds parameters which reduce the error between the measurement data and the model. The data is plotted as open circles in Fig. 6a) whereas the fitted function is plotted as lines in the same color code. An alternative way to evaluate if the receiver is saturating, is the explicit assumption of a square-root behavior from the photocurrent to the THz pump power. This is done with the responsivity in $\mu A/\sqrt{W}$ and displayed in Fig. 6b). The maximum responsivity is approximately 3 $\mu A/\sqrt{W}$ and has nearly no roll-off for higher THz powers. A receiver under strong saturation should show a strong decrease in responsivity. Fig. 6c) displays the fitted parameter $d$ with closed circles from Fig. 6a) for the different laser probe powers. The values have no clear trend and fluctuate between 0.2 and 0.42, which indicate some saturation of the receiver. The max. responsivity in Fig. 6d) shows a clear trend, with an optimum 3 $\mu A/\sqrt{W}$ reached for a laser probe power of 9 mW and confirms the optimal probe power for maximum DR in frequency domain.

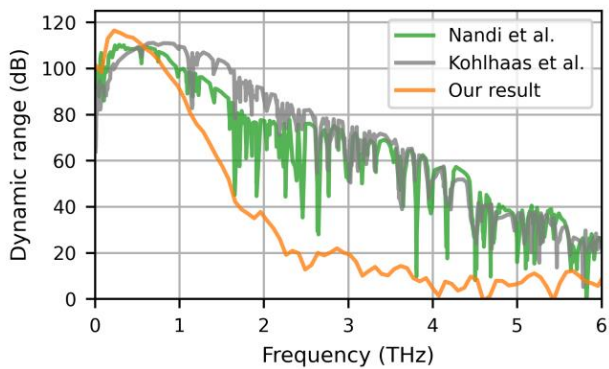

**Fig. 7** Comparison of our result (orange) under optimized conditions for maximum peak dynamic range in frequency domain (20.1 mW THz power, 6 mW laser probe power) to the recent, state-of-the-art peak dynamic-ranges achieved with PCA emitter and receiver by Nandi et al. (green) [12] and Kohlhaas et al. (grey) [11]. Extracted from [12] ©2023 IEEE and [11] ©2023 AIP.

Afterwards, a clear roll-over in terms of max. responsivity can be seen for higher probe powers, in strong contrast to the increase of THz power which did not show this behavior.

Fig. 7 shows the highest obtained DR, which we measured with the high-power emitter/low-noise receiver system (orange), resulting in 115 dB peak DR. The delay range is approximately 15 ps and the measurement time is 120 s. It is obtained with a THz power of 20.1 mW and a laser probe power 6 mW. The current record dynamic range of 110 dB by Nandi et al. [12] with 60 s measurement time (green) and 111 dB from Kohlhaas et al. [11] with 120 s measurement time (grey) is displayed as well. Due to longer delay range and higher bandwidth, water vapor absorption lines can be observed in the spectra of the other measurements displayed here.

V. CONCLUSION AND OUTLOOK

In summary, we present a combination of a high-power THz source based on the titled pulse front method in lithium niobate with a highly sensitive photoconductive receiver which is optimized for 1030 nm. We observed no significant saturation from the receiver even at highest THz powers of more than 20 mW. We noticed optimal laser probe powers to maximize the dynamic range in the time (frequency) domain of 6 mW (9 mW), where higher probe powers showed a clear roll-over. The novel detector is an excellent match to a high-power THz source and has similar performance as current, state-of-the-art results for 1550 nm [11], [12].

A clear next step is to improve the THz power even further [21], to make possible saturation/roll-over effects from too much THz power visible or, even better, reach higher dynamic ranges. Further improvements can be realized with active beam stabilization and stabilizing the oscillator. Implementing such stabilization schemes should lead to a lower noise-floor and increased dynamic range. From a bandwidth perspective, lithium niobate is, for a single-cycle THz source, rather narrowband and typical commercial systems present bandwidths up to 6 THz, even though their amplitude is strongly falling-off towards higher frequencies. Depending on the experiment, where i.e. a strong absorption at low THz frequencies needs to be probed, lithium niobate has a higher power spectral density, which can be better suited. We want to point out, that for applications calling for even broader bandwidths, two-color plasma sources [20] or organic crystals like BNA [59], [60] can exceed the bandwidth of typical photoconductive antennas and could serve as a powerful emitter source with a low-noise detection source. It should be noted, however, that shorter laser pulse durations are then necessary to reach such high bandwidths. We also highlight the future potential direction of large-area photoconductive emitters [15], [61], [62] using high average power lasers.

The reduced repetition rate of 13.4 MHz of our source is lower compared to the typically employed 100 MHz fiber lasers from commercial THz-TDS. The lower number of pulses is still more than sufficient to sample the electric field and in principle a much higher sampling rate (number of traces per second) could be achieved. The current limitation for our system is the speed of the mechanical delay line. Some commercial systems employ a second ultrafast laser to implement systems like asynchronously optical sampling (ASOPS) or ECOPS, to get thousands of traces per second [10]. Even though they increase the speed, they did not show (so far) new dynamic range records for similar recording times as those presented here. In the future, we expect higher repetition rate sources with similar field strength – i.e. – with higher average power to become available, facilitating shorter measurement times at these high DRs. Furthermore, future work on optimizing the noise performance of high average power lasers in this context will support these goals.

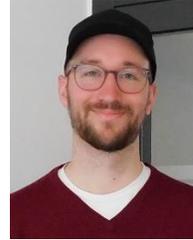

**Tim Vogel** received his B.Sc. and M.Sc. degree in physics from Heinrich-Heine University Düsseldorf, Germany. He wrote his master thesis at Airbus Defence & Space in Friedrichshafen, Germany. In 2019, he joined the Photonics and Ultrafast Laser Science (PULS) group at Ruhr-University Bochum, Germany. He is currently pursuing his PhD on high power THz generation and optimization of THz-Time Domain Spectroscopy.

Tim Vogel is member of the German Physical Society (Deutsche Physikalische Gesellschaft e.V.), European Physical Society (EPS), and Optica.

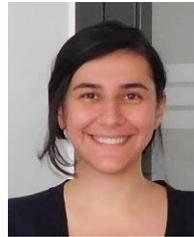

**Samira Mansourzadeh** received the M.Sc. degree in Laser and Photonics from Ruhr-Universität Bochum, Germany in 2018 where she worked on pulse laser compression and characterization. Since 2018, she has been pursuing her Ph.D. degree as a research assistant with the chair of Photonics and Ultrafast Laser Science (PULS). Her research interest includes the ultrafast laser developing, THz generation, detection, and its application in the imaging technology.

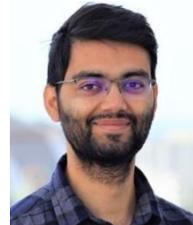

**Uttam Nandi** received the B.Tech. degree in electronics and communication engineering from the Indian Institute of Technology Guwahati, Guwahati, India, in 2014, the M.Sc. degree in wireless, photonics, and space engineering from the Chalmers University of Technology, Gothenburg, Sweden, in 2016, and the Ph.D. degree from the Technical University of Darmstadt, Darmstadt, Germany. In 2017, he joined the Terahertz Devices and Systems Laboratory, Technical University of Darmstadt. His research focuses on the development of terahertz pulsed receivers and emitters based on ErAs:In(Al)GaAs photoconductors.

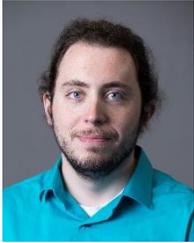
**Justin Norman** received B.S. degrees in chemical engineering and physics from the University of Arkansas at Fayetteville, Fayetteville, AR, USA in 2013. He received a Ph.D. in Materials from the University of California, Santa Barbara, CA, USA in 2018 as a National Science Foundation Graduate Research Fellow and a Frenkel Foundation Fellow. From 2018-2020, he continued as a postdoc at the University of California, Santa Barbara. He is currently at Quintessent, Inc.

His research interests have included the growth of InAs quantum dots via molecular beam epitaxy for applications in photonics and quantum electrodynamics, heteroepitaxy of III-V materials on Si for photonic integration, and other III-V based structures for optoelectronics. His current work is focused on the commercialization of quantum dot based optical communications systems for datacom and computing applications.

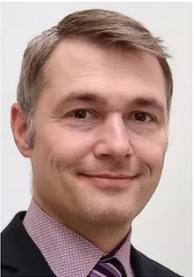
**Sascha Preu** (Member, IEEE) received the diploma degree in 2005, and the Ph.D. degree in physics (summa cum laude) from the Friedrich-Alexander University Erlangen-Nürnberg, Erlangen, Germany, in 2009. From 2004 to 2010, he was with the Max Planck Institute for the Science of Light, Erlangen. From 2010 to 2011 he was with the Materials Department and Physics Department, University of California, Santa Barbara, Santa Barbara, USA. From 2011 to 2014, he worked as the Chair of Applied Physics, University Erlangen-Nürnberg. He is currently a full Professor with the Department of Electrical Engineering and Information Technology, Technical University of Darmstadt, Germany, leading the Terahertz Devices and Systems Laboratory. His research interests include the development of semiconductor-based terahertz sources and detectors, including photomixers, photoconductors and field effect transistor rectifiers as well as terahertz systems constructed thereof. In 2017 he received an ERC starting grant for developing ultra-broadband, photonic terahertz signal analyzers. In 2022 he received an ERC Proof of Concept Grant for further developing photonic Terahertz spectrum analyzers. He also works on applications of terahertz radiation, in particular the characterization of novel terahertz components and materials.

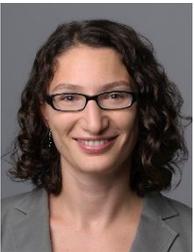
**Clara J. Saraceno** (Member, IEEE) was born in 1983 in Argentina. In 2007 she completed a Diploma in Engineering and an MSc at the Institut d'Optique Graduate School, Paris. She completed a PhD in Physics at ETH Zürich in 2012, for which she received the ETH Medal and the European Physical Society (Quantum Electronics and Optics Division) thesis prize in applied aspects in 2013. From 2013-2014, she worked as a Postdoctoral Fellow at the University of Neuchatel and ETH Zürich, followed by a postdoc position from 2015 – 2016 at ETH Zürich. In 2016, she received a Sofja Kovalevskaja Award of the Alexander von Humboldt Foundation and became Associate Professor of Photonics and Ultrafast Science in the Electrical Engineering Faculty at the Ruhr-University Bochum, Germany. In 2018 she received an ERC Starting Grant, and in 2019 was selected as an OSA Ambassador. Since 2020, she is a full professor at the Ruhr University in Bochum and her current main research topics of her group include high-power ultrafast lasers and Terahertz science and technology.